# Interleaved diffractive networks for information transfer through random diffusers

Yuhang Li[1,2], Yiyang Wu[1,2], Shiqi Chen[1,2], Xilin Yang[1,2], and Aydogan Ozcan[*,1,2,3]

[1]Electrical and Computer Engineering Department, University of California, Los Angeles, CA, 90095, USA.

[2]California NanoSystems Institute (CNSI), University of California, Los Angeles, CA, USA.

[3]Bioengineering Department, University of California, Los Angeles, 90095, USA.

[*]Correspondence: Aydogan Ozcan. Email: ozcan@ucla.edu

**Abstract**

Transferring optical information through random diffusers is a critical yet challenging task. In this work, we introduce a cascaded diffractive optical network for information transfer through random and unknown diffusers, achieved through a series of passive, structured layers physically interleaved within the scattering medium. These interleaved diffractive layers are optimized to mitigate the scattering process without requiring digital computing. The performance of this all-optical system was quantified as a function of several physical parameters, including the diffractive processor's depth, the physical layout of the diffractive layers, and the statistical properties of the scattering medium. To further enhance the performance and robustness of information transfer through a scattering medium, we also developed a hybrid (optical-digital) system that coupled the diffractive processor with a jointly-trained digital neural network, which was shown to achieve superior reconstruction fidelity even when the input object information was subjected to unknown random rotations, shifts, and scaling. We also experimentally validated this system using a fabricated multi-layer diffractive system in the visible spectrum, demonstrating reliable information recovery through random diffuser layers. Our numerical and experimental results demonstrate the capabilities of an interleaved diffractive processor architecture to recover optical information through volumetric diffusive media, which can find applications in biomedical imaging, telecommunications, and remote sensing.

**Introduction**

Imaging or transferring information through volumetric scattering media, such as biological tissue or atmospheric fog, represents a long-standing and significant challenge in optical sciences[1–6]. When light propagates through such complex media, it undergoes multiple scattering events that scramble the wavefront, destroying spatial information and collapsing a clear image into a seemingly random speckle pattern. Overcoming this fundamental limitation is critical for numerous applications, including deep-tissue imaging, remote sensing, and underwater navigation[1,2,7–10]. When the transmission matrix of the scattering medium is known, computational methods can, in principle, recover the original image. However, accurately and periodically measuring this evolving transmission matrix is often infeasible in practice[11]. Other approaches, such as wavefront shaping and adaptive optics[1,12–17], can provide powerful solutions but are often limited by their reliance on complex, active feedback systems, slow operational speeds, and limited fields-of-view. Various emerging computational methods that digitally process the resulting speckle patterns can also be used to recover the hidden information; however, these are inherently limited by the latency of data acquisition/transfer and digital processing, as well as the power consumption of digital computing[5,18–26], which can be prohibitive for real-time applications where low-latency and low-power information recovery is crucial.

Here, we present an interleaved diffractive optical network that leverages a series of passive, structured layers to all-optically transfer information through complex volumetric diffusers, as shown in **Fig. 1**. The trainable diffractive layers, optimized using deep learning, are physically interleaved within a random scattering medium. The information recovery process is performed purely in the optical domain through the propagation of the scattered fields within the diffractive processor. We demonstrated the effectiveness of this framework through various numerical models, systematically analyzing the impact of key physical parameters, including the depth of the diffractive network, the statistical properties of the random scattering medium, and the physical layout of the diffractive layers. Numerical blind testing results validated the feasibility of our approach, with the diffractive network consistently recovering arbitrarily selected unseen object information that was obscured by random and unknown volumetric diffusers – never seen before. Furthermore, we designed a hybrid optical-digital system for optical information transfer through a volumetric diffuser with enhanced performance, and evaluated its resilience under various conditions, including random rotations, shifts, and scaling of the input objects, confirming the robust performance of the hybrid model against such unknown variations at the input. In addition to numerical analyses, we experimentally verified this approach using a fabricated multi-layer diffractive system for visible light, successfully demonstrating information recovery through random diffusers. We believe that these interleaved diffractive networks for information transfer through scattering media might find various applications in biomedical imaging, remote sensing, and free-space communications.

**Results**

**Imaging through diffusers with interleaved diffractive networks**

**Figure 1a** illustrates the schematic of an interleaved diffractive optical network trained to reconstruct an object obscured by a volumetric diffuser. To model random volumetric diffusers, we used a cascade of $M$ thin diffusers with random and independent phase profiles, separated by a fixed axial distance. Between each diffuser plane, we inserted $K$ diffractive layers, aiming to incrementally correct the scattered wavefront as it propagates through the scattering medium. During the training phase, handwritten digits were used as amplitude input objects[27], and the optical field propagated through the diffusers and subsequent diffractive layers, forming an intensity pattern at the output plane. To enhance the network's generalization capability for imaging through complex volumetric media, we introduced $n$ random diffusers during each training epoch (**Fig. 1b**). The diffractive network was designed and trained numerically using the angular spectrum method for free-space propagation; see the **Methods** section. The phase values of the features at each diffractive layer were iteratively updated via error backpropagation, maximizing the Pearson Correlation Coefficient (PCC) between the network's output intensity and the ground truth target image (see the **Methods** section for details). The training was concluded after 100 epochs, during which the interleaved diffractive processor was exposed to a total of $N = 100n$ unique volumetric diffuser realizations. Following this one-time training process, the resulting interleaved diffractive processor was fixed and subjected to blind testing. In this testing phase, the network's performance was evaluated using previously unseen input objects from the MNIST test set, which were obscured by entirely new, unseen volumetric random diffusers. **Figure 1c** displays the optimized phase profiles of the diffractive layers. The trained layers exhibit complex patterns, including densely-packed microlens-like phase structures, which contribute to image formation, while the surrounding spatial phase features help reduce random diffuser-induced spatial distortions by coupling their energy outside of the output field-of-view[28]. Collectively, these optimized layers function to unscramble the light distorted by the random diffusers, helping to form clearer images of the input objects at the output plane.

**Figure 2** compares two diffractive architectures, each trained from scratch, for imaging through a volumetric diffuser: the interleaved diffractive network that placed diffractive layers inside the scattering volume after each diffuser unit (**Fig. 2a**), and a non-interleaved baseline design in which all the diffractive layers were placed outside the diffuser volume (**Fig. 2b**). **Figure 2c** shows representative test results for various handwritten digits and grating targets. It is worth noting that such grating targets were not included during the training and serve as an external test set to showcase the network's generalization to new types of objects. The interleaved architecture consistently preserves the global structure of the objects across unseen/new random diffusers, producing sharper digit strokes and cleaner separation of gratings (see **Movie S1** for additional results). In contrast, concentrating the diffractive layers outside the random diffuser volume led to pronounced speckle patterns and noticeable structural loss. For reference, propagation through volumetric diffusers without any diffractive layers resulted in stronger speckle distortion, as illustrated in **Fig. 2c**. These results indicate that distributing the optimized diffractive layers within the scattering volume enables progressive distortion correction after each random diffuser and markedly improves the robustness to random diffuser variability.

**Impact of the number of optimized diffractive layers**

We investigated the impact of diffractive network's depth on the imaging performance through volumetric diffusers by varying the number of diffractive layers, $K$, within each of the diffuser units, while keeping the axial distance between the random diffusers constant (**Fig. 3a**). The quantitative results, shown in **Fig. 3b**, reveal that the image reconstruction fidelity monotonically improved as $K$ increased from 2 to 5 diffractive layers for both known and unseen/new test diffusers; here, known diffusers refer to those encountered during the training and unseen diffusers were never seen before. The comparable performance achieved on new volumetric diffusers demonstrates the diffractive network's generalization capability, indicating that it learned the underlying physical principles governing light propagation through scattering media, rather than merely memorizing specific examples. Furthermore, the most significant performance gain was observed when increasing $K$ from 2 to 3. These findings are also visualized in **Fig. 3c**. Without the diffractive layers, the outputs were unrecognizable speckle patterns. With two diffractive layers per unit, the global shapes of the digits began to emerge, albeit with some blurring. Further increasing the number of diffractive layers per diffuser unit to $K = 4$ and $K = 5$ enhanced finer features, such as stroke continuity and contrast, confirming the improvements seen in the all-optically reconstructed data. Collectively, these results indicate that increasing the model capacity[29,30] strengthens the network's robustness against volumetric random scattering.

**Impact of diffuser depth and statistics**

The performance of the diffractive processor is also dependent on the characteristics of the random diffusers. We evaluated this relationship from two perspectives: the overall thickness of the random scattering medium, which was modeled by the number of diffuser units $M$, and the diffuser's internal statistical properties, defined by the phase correlation length, $C$. First, we evaluated the network's performance against increasing scattering thickness by changing the number of diffuser units, $M$, as illustrated in **Fig. 4a**. The quantitative results, reported in **Fig. 4b**, reveal a trade-off between the stronger distortion from thicker volumetric diffusers and the image reconstruction capability of deeper diffractive processors. Initially, the reconstruction performance slightly increased as $M$ increased from 2 to 3, which can be attributed to the corresponding increase in the total number of optimized diffractive layers within the volume. However, as $M$ was increased further, image reconstruction fidelity degraded. This indicates that beyond a certain point, the distortion introduced by additional random scattering overwhelmed the corrective capacity of the added diffractive layers. This trend is also visualized in **Fig. 4c**, which displays representative output images for different numbers of diffuser units. While the diffractive processor successfully reconstructed the input test digits through $M = 1$ or 2 units, the image quality progressively deteriorated with three and four units, exhibiting increased noise and a loss of features. The outputs without diffractive layers also confirmed this escalating challenge: a rough profile of the digit was discernible for $M = 1$, but the outputs became unrecognizable speckle patterns for $M \geq 2$.

Next, we investigated how the diffuser's spatial characteristics impact the optical image reconstruction performance. We trained and tested the diffractive network architecture ($M =$ 2, $K = 2$) against diffusers with varying correlation lengths, $C$, as shown in **Fig. 5a**. The parameter $C$ approximately defines the average "grain size" of the scattering features, where a larger correlation length corresponds to a weaker scattering effect and a less challenging image reconstruction problem (see the **Methods** section for details). **Figure 5b** confirms this relationship, showing a monotonic increase in the output image PCC as the diffuser correlation length increases from 7.5 $\lambda$ to 15 $\lambda$. This improvement in reconstruction fidelity was consistent for both known and new/unseen random diffusers. We also tested diffractive networks (trained with different correlation lengths) by removing random diffusers, as shown in green curves in **Fig. 5b**. Notably, models trained with weaker diffusers (larger $C$) also achieved higher performance in this test. This suggests a trade-off during optimization: when the volumetric scattering was weak, more of the diffractive network's degrees of freedom are allocated to the primary image reconstruction task, whereas for stronger scattering, more capacity must be dedicated to the filtering of distortions. Visualizations of some test object reconstructions are shown in **Fig. 5c**. For a smaller correlation length (e.g., $7.5\lambda$), despite the strong volumetric scattering, our system can still reconstruct the underlying shape of the digits, albeit with some noise. In contrast, for larger correlation lengths (e.g., 10 $\lambda$ and 12.5 $\lambda$), the diffractive processor produced significantly clearer and more accurate images of the target objects.

**Impact of axial separation between successive diffractive layers**

We also investigated the influence of the axial separation, $d$, between consecutive planes (including both the diffractive layers and diffusers) on the network's overall performance. Using the architecture shown in **Fig. 6a**, we trained and tested several diffractive processors where the inter-layer distance was varied. The quantitative results, plotted in **Fig. 6b**, reveal that the image reconstruction fidelity was sensitive to this distance. For both known and new random diffusers, the output PCC was highest at the smallest separation ($6.7\lambda$), showing a significant drop at intermediate distances before plateauing for separations beyond $20\lambda$. This performance trend is visually substantiated by the image reconstruction results presented in **Fig. 6c**. The images produced by the most compact network ($d = 6.7\lambda$) were the sharpest and of the highest image quality. As the layers were moved farther apart, the output images became progressively noisier and lost structural details, consistent with the initial degradation seen in the quantitative results.

These findings suggest that a more compact diffractive network with a larger effective numerical aperture between the optimized layers is more effective at manipulating the optical wavefront to correct for random distortions caused by volumetric scattering media.

**Hybrid optical-digital system for enhanced image reconstruction performance**

To further improve the fidelity of information transfer through volumetric diffusers, especially for highly scattering media, we also implemented a hybrid optical-digital system. As illustrated in the schematic in **Fig. 7a**, this architecture utilizes an all-optical diffractive processor to perform analog pre-processing of the scattered light, thereby forming an intermediate image at the camera plane. The resulting pre-processed image is then fed into a jointly-trained digital neural network for the final digital processing and image reconstruction step. Both the diffractive layers and the digital network were co-optimized through end-to-end backpropagation to maximize the PCC between the final reconstructed output and the target information.

To demonstrate the effectiveness of this hybrid approach, we considered a challenging configuration with $M = 4$ random diffuser units, where the all-optical diffractive solution alone fails to produce a clear image at its output plane. As shown in **Fig. 7b** and **Movie S2**, the results highlight the advantages of the hybrid approach over all-optical diffractive designs. For the hybrid model, where the diffractive optical network and digital neural network were trained jointly, the all-optical output $O_{Optical}$ at the camera plane appeared blurry; however, the final hybrid output $O_{Digital}$ was substantially cleaner, demonstrating the digital network's powerful ability to denoise the optically processed signals.

Furthermore, we evaluated the system's robustness against input perturbations by applying random rotations, shifts, and scaling to input digits that were never seen before. Even when faced with these "attacked" inputs, the hybrid system successfully recovered the core information and produced a high-fidelity image reconstruction; also see **Movie S2**. This synergy between analog and digital processing creates a robust and highly effective channel for transferring information through complex, scattering media.

**Experimental validation of interleaved diffractive networks**

To experimentally validate the interleaved diffractive network for information transfer through random diffusers, we implemented and trained a two-layer diffractive system interleaved with two random phase diffusers; see **Fig. 8a**. During the in silico training stage, the optical diffractive layers and the digital neural network backend were jointly optimized using thousands of different random phase diffusers using MNIST handwritten digit images as the training dataset[27]. In this hybrid training framework, the input digit images were represented as phase-only objects spanning the range $[0, \pi]$, while the corresponding target images were normalized to [0,1]. Accordingly, the hybrid system learned a phase-to-intensity transformation, which is a nonlinear transformation. After convergence, the entire structure, including both the optimized diffractive phase profiles and new random phase diffusers (never used during the training stage), was fabricated using two-photon polymerization lithography, forming a monolithic four-layer stack (see **Methods** for details).

The experimental setup and the 3D-fabricated multi-layer structure are shown in **Fig. 8b**, including a photograph of the optical system and a schematic of the interleaved layers. The fabricated structures and their corresponding numerical phase profiles are presented in **Fig. 8c**, which also reports microscopic images of the random phase diffusers and optimized diffractive layers alongside their numerical phase distributions. During blind testing, the input object test patterns were displayed on a spatial light modulator (SLM) and optically projected onto the fabricated four-layer structure. The resulting output intensity distribution was captured by a camera. The fabricated stack consisted of two optimized diffractive layers interleaved with two random, newly generated phase diffusers that were never used during training. The structure was kept physically fixed throughout the experiments, and generalization was assessed using previously unseen input objects.

In total, 2000 experimental measurements were acquired under these fixed scattering conditions. **Figure 8d** presents a direct comparison between the numerical predictions and the experimentally measured images at the output CMOS image sensor. Despite careful fabrication and alignment, discrepancies between simulation and experiment were observed due to model mismatches, fabrication tolerances, and imperfections. When the pre-trained digital backend was directly applied to the experimental measurements without further adaptation, the reconstruction quality degraded noticeably, highlighting the gap between the numerical training and physical implementation. This performance drop motivates the need for transfer learning. Accordingly, the 2,000 experimental measurements were partitioned into training and test subsets with varying test split ratios, where the test split ratio denotes the fraction of the full dataset held out for testing and excluded from parameter updates, while the remaining data were used for training. For each split configuration, the digital backend underwent in situ transfer learning using the corresponding training subset, while the optical diffractive layers remained physically unchanged. The reconstruction performance was then evaluated on the held-out test set to quantify generalization.

**Figure 8e** shows the test loss as a function of the test split ratio. Across all the tested split ratios $(0.05 - 0.9)$, the interleaved diffractive network consistently achieved low test loss and stable imaging performance under different levels of in situ training data. As the amount of in situ training data decreased, the test loss gradually increases, reflecting the reduced ability of the digital backend to adapt to experimental deviations using limited data. These findings are further corroborated by the reconstructed examples shown in **Fig. 8f**. Across different split ratios, the reconstructed images closely matched the target images and preserved the complete object structure. When the training subset becomes smaller, the reconstruction quality degrades gradually, with more noticeable structural distortions appearing in the recovered images. Collectively, these results experimentally demonstrate the effectiveness of the optimized interleaved diffractive layers, which provide structured compensation for the random scattering process.

**Discussion**

In this work, we introduced interleaved diffractive processors for high-fidelity information transfer through random volumetric diffusers. By interleaving optimized layers within a random scattering medium, the passive, all-optical processor learns to invert the effects of multiple scattering and recover the hidden information. This interleaved architecture enables the optical network to successively correct the optical wavefront during propagation, rather than attempting to digitally reconstruct a highly distorted field at the output plane. Additionally, the system's performance and limitations were systematically characterized against different physical parameters, including the network's depth, the physical spacing of its layers, and the statistical properties of the random scattering medium. We further demonstrated that by integrating the all-optical processor with a jointly trained digital neural network, the system's performance and robustness to input variations can be significantly enhanced, resulting in a powerful hybrid information recovery system. Furthermore, beyond numerical validation, we experimentally demonstrated the concept using a fabricated multi-layer diffractive system. The measured results confirmed reliable information recovery under complex random scattering conditions. These experimental findings substantiate the physical feasibility of interleaved optimized diffractive layers within scattering media to achieve structured optical compensation.

While the proposed framework demonstrates robust imaging through volumetric diffusers, several challenges remain. Fabricating multi-layer diffractive systems with precise axial alignment poses practical difficulties, especially as the number of interleaved layers increases. These alignment challenges may be mitigated through a vaccination strategy, in which misalignment perturbations are injected during training to help the system develop robustness against such errors[31]. Furthermore, scaling the approach to larger apertures or broadband operation would be another challenging task. These spectral effects can be explicitly incorporated into the training loss function, enabling the network to jointly optimize performance across multiple wavelengths[32,33]. On the computational side, training with more realistic scattering models or experimentally measured diffusers may further enhance generalization to practical scenarios, albeit at the cost of increased optimization complexity.

Looking ahead, the presented interleaved diffractive processor framework holds promise for applications in compact imaging systems, endoscopic or through-tissue imaging, and vision through dynamic scattering media. It may also be extended to multispectral or polarization-sensitive modalities, or combined with hybrid architectures for real-time, energy-efficient optical information processing.

## Methods
### Neural Network Architecture for the Hybrid Optical-Digital System

We employed a compact three-level U-Net architecture with a symmetric encoder–decoder structure and skip connections between the corresponding layers for the hybrid optical-digital system reported in **Figs. 7-8**. Each convolutional block consisted of two $3 \times 3$ convolutional layers followed by ReLU activation functions, while $2 \times 2$ max-pooling and transposed convolutions were used for down- and up-sampling, respectively. Dropout ($p = 0.2$) was applied in intermediate blocks to improve generalization. The encoder progressively increased the number of feature channels (2, 4, 8) while reducing the spatial resolution by a factor of two at each stage. The bottleneck block expanded the feature dimension to 16 channels to capture global context. The decoder mirrored the encoder, restoring the spatial resolution through learned upsampling and concatenating skip features to recover fine details. A final $1 \times 1$ convolution followed by a sigmoid activation produced the single-channel output $O_{Digital}$. This lightweight network contained approximately 7.6k trainable parameters. All weights were initialized using the Kaiming initialization method.

**Training Objective and Evaluation**

The interleaved diffractive network was trained to reconstruct target images $I_{target}$ at the sensor plane. The loss function was defined as:

$$Loss(O, I_{target}) = 1 - \text{PCC}(O, I_{target})$$

Where PCC was calculated using:

$$\text{PCC}(O, I_{target}) = \frac{\sum(O(x,y) - \overline{O}) \cdot (I_{target}(x,y) - \overline{I_{target}})}{\sqrt{\sum(O(x,y) - \overline{O})^2 \cdot (I_{target}(x,y) - \overline{I_{target}})^2}}$$

where $\overline{O}$ and $\overline{I_{target}}$ were the average intensity values of the output $O$ and the target image $I_{target}$, respectively.

**Training Implementation**

Numerical simulations of interleaved diffractive optical networks with volumetric diffusers were performed using coherent monochromatic illumination at a wavelength of $\lambda = 0.75\ mm$. The system consisted of $K$ trainable diffractive layers interleaved with $M$ random phase diffusers, with axial separations set to $d$ between consecutive planes. During the training, input images were grayscale and resized to $160 \times 160$ pixels ($64\lambda \times 64\lambda$) before being fed into the network, following the wave propagation steps described earlier. Each diffractive layer was composed of $240 \times 240$ trainable phase features ($96\lambda \times 96\lambda$), with a lateral feature size of ~$0.5\lambda$. The central $160 \times 160$ region ($64\lambda \times 64\lambda$) of the sensor plane was used for the loss function calculation.

Experimental training of the hybrid interleaved diffractive system was performed under coherent monochromatic illumination at $\lambda = 635$ nm. The fabricated four-layer stack consisted of two optimized diffractive layers interleaved with two random phase diffusers (never used during the training process). Each trainable diffractive layer comprised $512 \times 512$ phase features with a lateral pixel size of 4 μm.

During data acquisition, input object patterns of size $256 \times 256$ were displayed on an SLM (Pluto, 8 μm pitch size, HOLOEYE Photonics AG) and projected onto the fabricated structure. The input phase images were encoded over the range $[0, \pi]$, and the corresponding target images were intensity patterns normalized to [0,1]. The resulting output intensity distributions were recorded by a CMOS camera at the sensor plane. Since the experimental task mapped phase-only input objects into target intensity images, the system implemented a phase-to-intensity transformation, corresponding to an inherently nonlinear mapping. A total of 2,000 input object patterns were used, resulting in an experimental dataset acquired under fixed scattering conditions. The captured intensity images were cropped to $400 \times 400$ pixels before being fed into the digital neural network backend. For each train-test split configuration, the digital backend was initialized from the numerically pre-trained model and then further optimized using the training subset of the corresponding experimental dataset, i.e., via in situ transfer learning rather than training from scratch, while the optical diffractive layers remained physically fixed. The loss function was defined as $1 - PCC$ between the digitally reconstructed output images and the corresponding ground truth target images.

The interleaved networks were trained using Python (v3.10.6) and PyTorch (v1.12, Meta Platforms Inc.), with the Adam optimizer and a learning rate of $1 \times 10^{-3}$. For the hybrid model, the same settings were used except that the U-Net was trained with a learning rate of $1 \times 10^{-4}$. The batch size was set to 32, and training was run for 100 epochs and $n = 150$. Diffractive layer phase values were initialized to zero. The training was performed on a GeForce GTX 3090 GPU (Nvidia Corp.), and each training run typically required ~6 hours for numerical models and ~20 hours for experimental models, depending on network depth and number of diffuser units.

**Nano-fabrication process**

The interleaved diffractive network and random phase diffusers were fabricated using two-photon polymerization lithography (Quantum X shape, Nanoscribe GmbH). The fabrication procedure consisted of precision printing followed by chemical post-processing. The diffractive layers were quantized to 4-bit phase depth prior to fabrication. A commercial photoresist IP-DIP2 (Nanoscribe GmbH) and a $63 \times$ objective lens ($NA = 1.4$) were used to achieve a lateral feature size of ~2 $\mu m$.

Each diffractive layer required approximately 8 hours of printing time.

All diffractive layers and random phase diffusers were sequentially printed on the same 2-inch wafer (2" JGS2 Fused Silica Wafer, MSE Supplies). The fabrication process began with printing anchoring Fresnel lenses on the surface of a 2-inch wafer to provide mechanical support and alignment reference. After printing each layer or phase diffuser, the objective was translated laterally to allow excess resin to flow away from the printed structure. Residual resin surrounding the structure was carefully removed using a swab, while keeping the substrate holder fixed to maintain positional accuracy.

To construct the multi-layer structure, two thin spacers were prepared from a 170 $\mu m$-thick glass coverslip and coated with epoxy adhesive on both sides. These spacers were placed symmetrically on the left and right sides of the printed structure to serve as vertical supports for the subsequent layer. A new coverslip was then placed on top to define the interlayer spacing. Each adhesive interface contributed ~30 − 50 μm thickness. The total axial separation between each diffractive layer and its adjacent phase diffuser was calculated as (40 μm × 2 + 170 μm × 2), accounting for the epoxy adhesive layers, glass spacer thickness and the coverslips on which the layers were printed. Considering the refractive index of the coverslip glass ($n \approx 1.5$), the effective optical path length between the layers was estimated as 40μm × 2 + 170 μm + 1.5 × 170 μm. This distance was incorporated into the numerical training model to ensure consistency between simulation and fabrication. Following the spacer placement, fresh resin was added, and the system was returned to the previously recorded lateral coordinates. Autofocus was performed along the z-axis before initiating the next printing cycle. This procedure was repeated three times to complete the four-layer interleaved diffractive structure with random phase diffusers.

After fabrication, the sample underwent a sequence of solvent baths to remove unpolymerized resin and ensure structural integrity. The sample was first immersed in propylene glycol monomethyl ether acetate (PGMEA) for 5–10 min to dissolve residual photoresist. It was then transferred sequentially into an isopropyl alcohol (IPA) bath for 5 min and a Novec 7100 engineered fluid bath (methoxy-nonafluorobutane, HFE-7100) for an additional 5 min. The final solvent enabled rapid drying without nitrogen blow-drying, minimizing mechanical stress on the multilayer structure.

**Supplementary Information**: Forward Model of Interleaved Diffractive Networks through

**References**

1. Ji, N., Milkie, D. E. & Betzig, E. Adaptive optics via pupil segmentation for high-resolution imaging in biological tissues. *Nat Methods* **7**, 141–147 (2010).

2. Jang, M. *et al.* Relation between speckle decorrelation and optical phase conjugation (OPC)-based turbidity suppression through dynamic scattering media: a study on in vivo mouse skin. *Biomed. Opt. Express* **6**, 72 (2015).

3. Tan, R. T. Visibility in bad weather from a single image. in *2008 IEEE Conference on Computer Vision and Pattern Recognition* 1–8 (2008). doi:10.1109/CVPR.2008.4587643.

4. Ntziachristos, V. Going deeper than microscopy: the optical imaging frontier in biology. *Nat Methods* **7**, 603–614 (2010).

5. Bertolotti, J. *et al.* Non-invasive imaging through opaque scattering layers. *Nature* **491**, 232–234 (2012).

6. Horisaki, R., Okamoto, Y. & Tanida, J. Single-shot noninvasive three-dimensional imaging through scattering media. *Opt. Lett., OL* **44**, 4032–4035 (2019).

7. LOPES, A., NEZRY, E., TOUZI, R. & LAUR, H. Structure detection and statistical adaptive speckle filtering in SAR images. *International Journal of Remote Sensing* **14**, 1735–1758 (1993).

8. Bucher, E. A. Computer Simulation of Light Pulse Propagation for Communication Through Thick Clouds. *Appl. Opt., AO* **12**, 2391–2400 (1973).

9. Jaffe, J. S., Moore, K. D., McLean, J. & Strand, M. P. Underwater Optical Imaging: Status and Prospects. *Oceanography* **14**, 64–75 (2015).

10. Schettini, R. & Corchs, S. Underwater Image Processing: State of the Art of Restoration and Image Enhancement Methods. *EURASIP J. Adv. Signal Process.* **2010**, 1–14 (2010).

11. Kim, M., Choi, W., Choi, Y., Yoon, C. & Choi, W. Transmission matrix of a scattering medium and its applications in biophotonics. *Opt. Express* **23**, 12648 (2015).

12. Wang, K. *et al.* Direct wavefront sensing for high-resolution in vivo imaging in scattering tissue. *Nat Commun* **6**, 7276 (2015).

13. Li, J. *et al.* Conjugate adaptive optics in widefield microscopy with an extended-source wavefront sensor. *Optica, OPTICA* **2**, 682–688 (2015).

14. Vellekoop, I. M. & Mosk, A. P. Focusing coherent light through opaque strongly scattering media. *Opt. Lett., OL* **32**, 2309–2311 (2007).

15. Horstmeyer, R., Ruan, H. & Yang, C. Guidestar-assisted wavefront-shaping methods for focusing light into biological tissue. *Nature Photon* **9**, 563–571 (2015).

16. Nixon, M. *et al.* Real-time wavefront shaping through scattering media by all-optical feedback. *Nature Photon* **7**, 919–924 (2013).

17. Yang, J. *et al.* Anti-scattering light focusing by fast wavefront shaping based on multi-pixel encoded digital-micromirror device. *Light: Science & Applications* **10**, 149 (2021).

18. Feng, S., Kane, C., Lee, P. A. & Stone, A. D. Correlations and Fluctuations of Coherent Wave Transmission through Disordered Media. *Phys. Rev. Lett.* **61**, 834–837 (1988).

19. Edrei, E. & Scarcelli, G. Memory-effect based deconvolution microscopy for super-resolution imaging through scattering media. *Sci Rep* **6**, 33558 (2016).

20. He, H., Guan, Y. & Zhou, J. Image restoration through thin turbid layers by correlation with a known object. *Opt. Express, OE* **21**, 12539–12545 (2013).

21. Qiao, H. *et al.* GPU-based deep convolutional neural network for tomographic phase microscopy with $\ell 1$ fitting and regularization. *J Biomed Opt* **23**, 1–7 (2018).

22. Barbastathis, G., Ozcan, A. & Situ, G. On the use of deep learning for computational imaging. *Optica, OPTICA* **6**, 921–943 (2019).

23. Li, S., Deng, M., Lee, J., Sinha, A. & Barbastathis, G. Imaging through glass diffusers using densely connected convolutional networks. *Optica* **5**, 803 (2018).

24. Jiang, S., Guo, K., Liao, J. & Zheng, G. Solving Fourier ptychographic imaging problems via neural network modeling and TensorFlow. *Biomed. Opt. Express, BOE* **9**, 3306–3319 (2018).

25. Li, Y., Xue, Y. & Tian, L. Deep speckle correlation: a deep learning approach toward scalable imaging through scattering media. *Optica* **5**, 1181 (2018).

26. Yang, M. *et al.* Deep hybrid scattering image learning. *J. Phys. D: Appl. Phys.* **52**, 115105 (2019).

27. Deng, L. The MNIST Database of Handwritten Digit Images for Machine Learning Research [Best of the Web]. *IEEE Signal Processing Magazine* **29**, 141–142 (2012).

28. Luo, Y. *et al.* Computational imaging without a computer: seeing through random diffusers at the speed of light. *eLight* **2**, 4 (2022).

29. Kulce, O., Mengu, D., Rivenson, Y. & Ozcan, A. All-optical information-processing capacity of diffractive surfaces. *Light Sci Appl* **10**, 25 (2021).

30. Kulce, O., Mengu, D., Rivenson, Y. & Ozcan, A. All-optical synthesis of an arbitrary linear transformation using diffractive surfaces. *Light Sci Appl* **10**, 196 (2021).

31. Mengu, D. *et al.* Misalignment resilient diffractive optical networks. *Nanophotonics* **9**, 4207–4219 (2020).

32. Shen, C.-Y. *et al.* Broadband unidirectional visible imaging using wafer-scale nano-fabrication of multi-layer diffractive optical processors. *Light Sci Appl* **14**, 267 (2025).

33. Shen, C.-Y. *et al.* Multiplane quantitative phase imaging using a wavelength-multiplexed diffractive optical processor. *AP* **6**, 056003 (2024).

34. Lin, X. *et al.* All-optical machine learning using diffractive deep neural networks. *Science* **361**, 1004–1008 (2018).

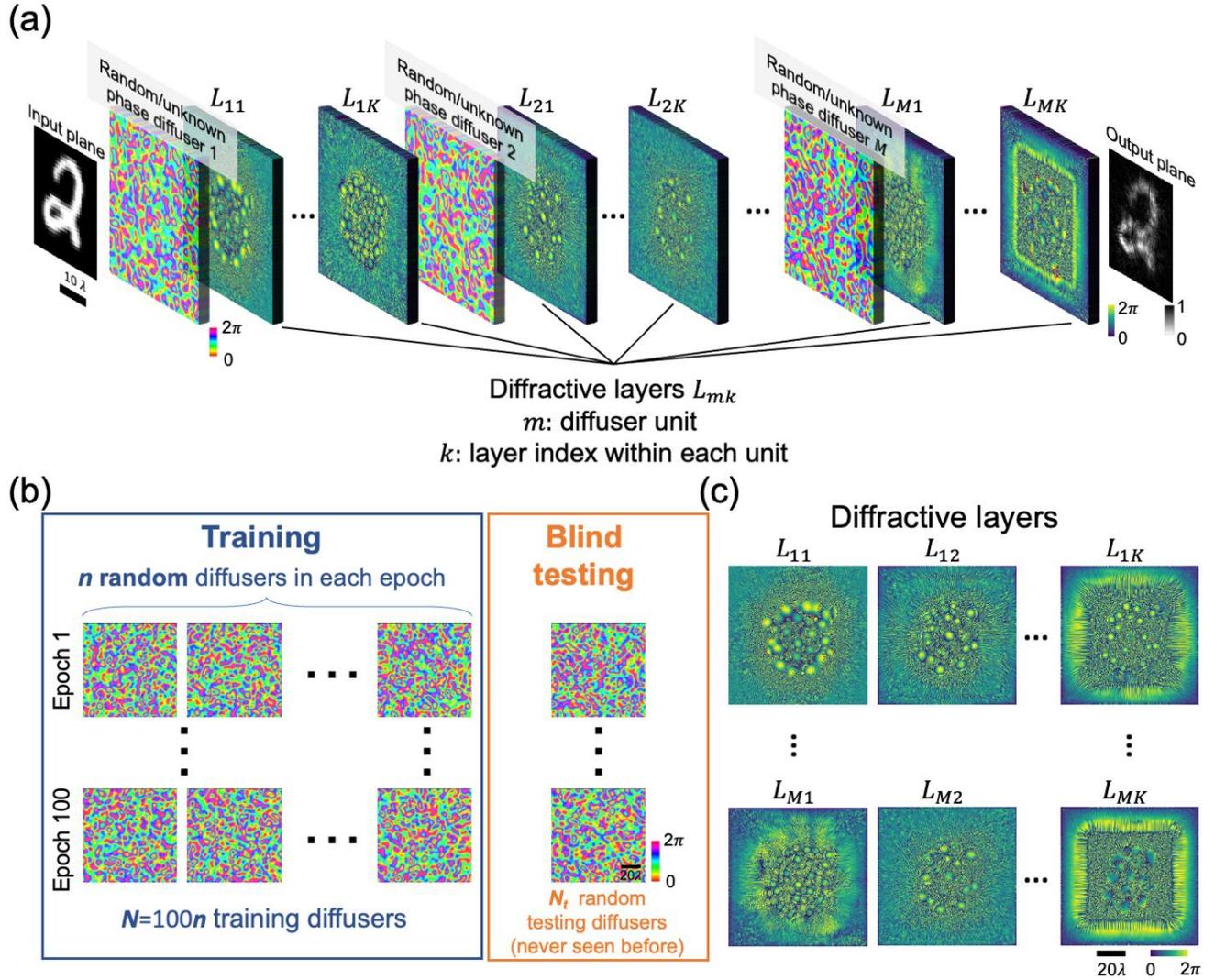

**Figure 1. Design and training of an interleaved diffractive network for imaging through a volumetric diffuser. (a)** Conceptual schematic of the diffractive imaging system. A volumetric diffuser, which scrambles the optical wavefront from an input object, is modeled by *M* separate random and unknown diffuser planes. Between each plane, *K* trainable diffractive layers are placed to progressively correct the distortions. **(b)** Training and blind testing strategy of the presented interleaved diffractive network. **(c)** Optimized phase profiles of the diffractive layers.

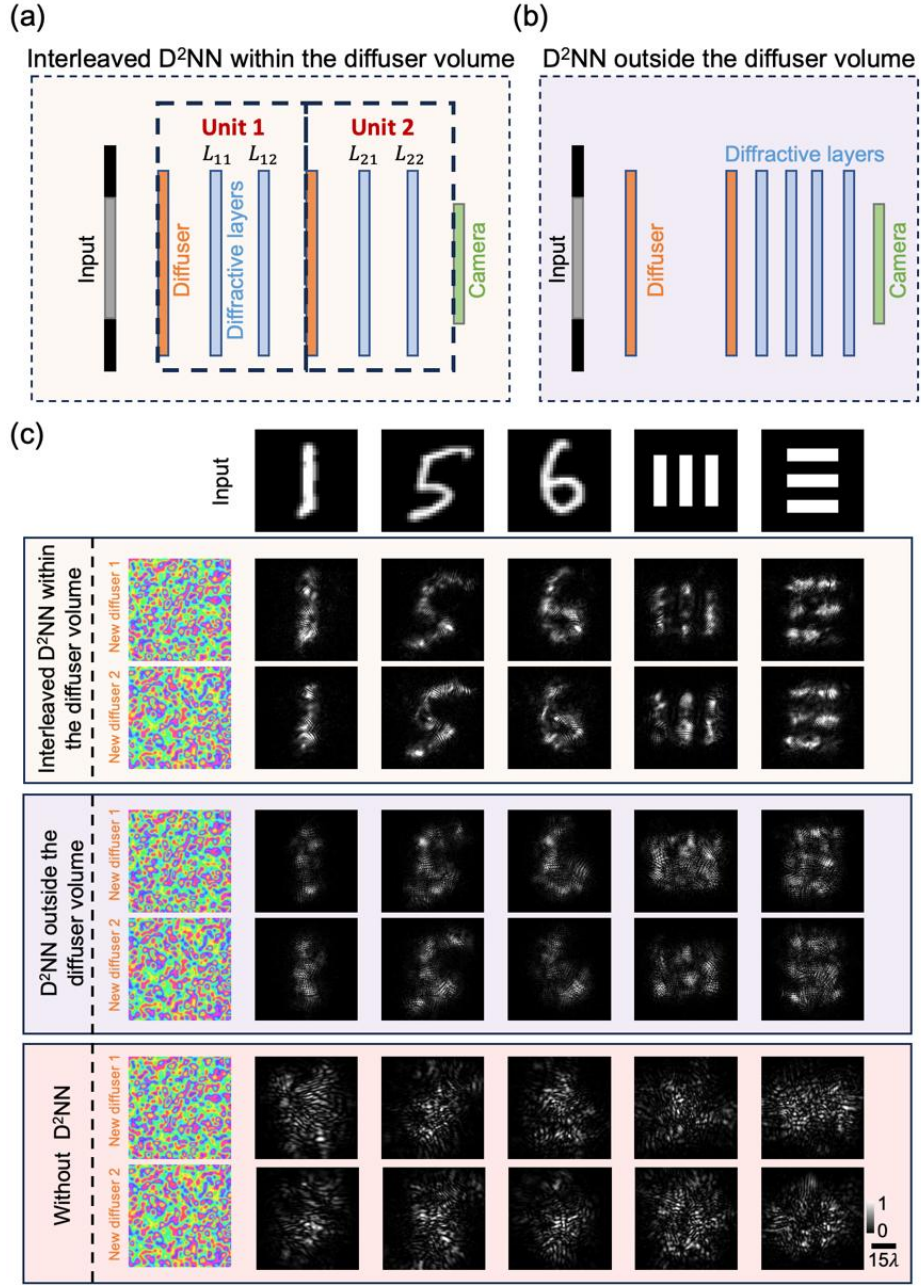

**Figure 2. Comparison of interleaved and non-interleaved diffractive network architectures for imaging through a volumetric random diffuser.** The schematic of **(a)** an interleaved diffractive network (D²NN) within the diffuser volume, and **(b)** a diffractive network outside the diffuser volume. **(c)** The all-optical reconstruction results for handwritten digits and grating targets imaged through new, random diffusers, comparing the performance of interleaved, non-interleaved, and no-diffractive-network-based models.

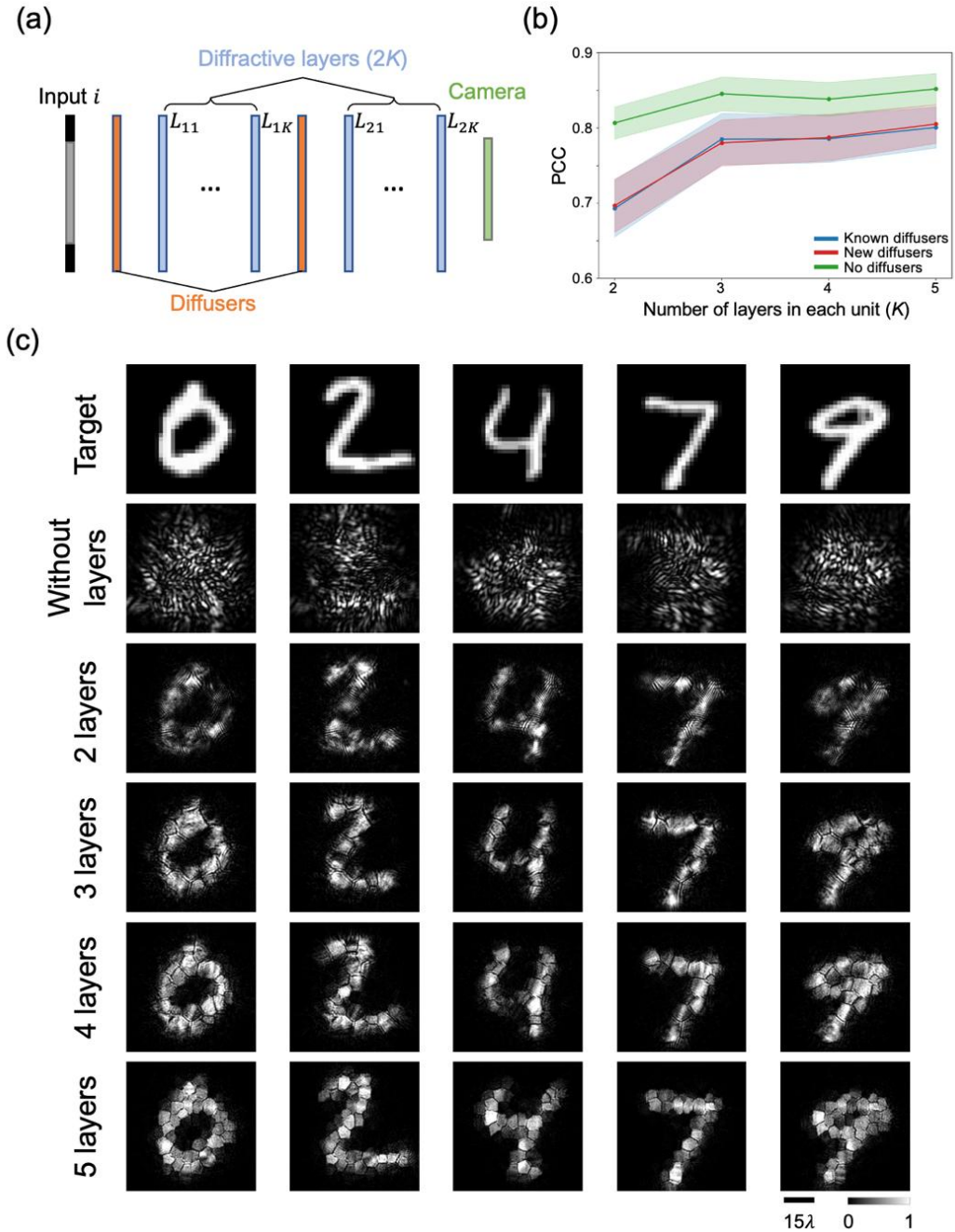

**Figure 3. Additional trainable diffractive layers improve the imaging performance through a volumetric random diffuser.** **(a)** Schematic of the interleaved diffractive network with two diffuser units, where the number ($K$) of diffractive layers in each unit was varied. **(b)** The PCC values calculated with respect to the ground truth objects as a function of the number of diffractive layers, $K$. Known diffusers refer to the diffusers that were used during the training stage, whereas new diffusers were never seen during training. **(c)** Test results of new objects through new random diffusers without diffractive layers and with $K = 2 - 5$ diffractive layers.

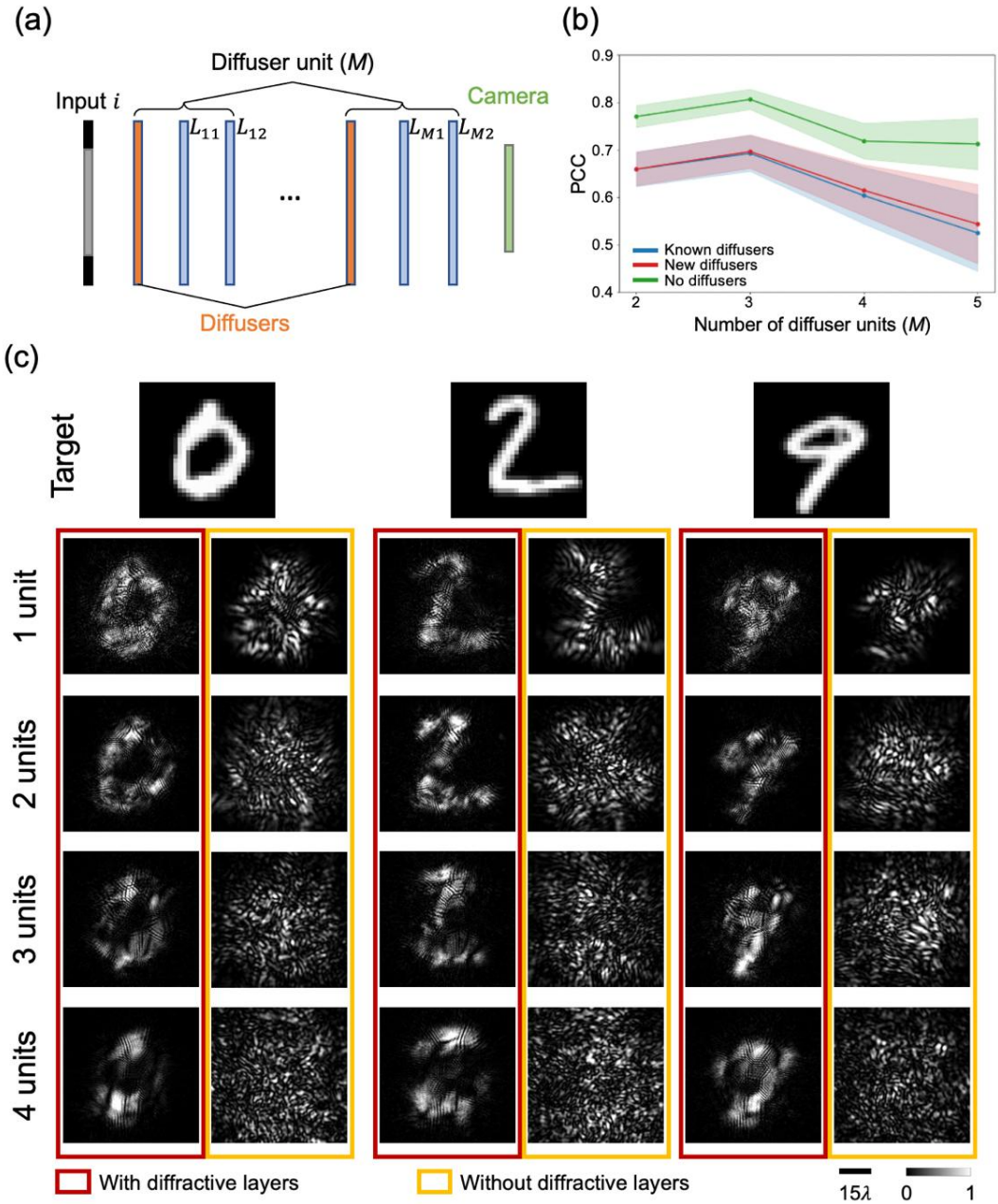

**Figure 4. Impact of the number of independent diffuser units on the imaging performance.** **(a)** Schematic of the interleaved architecture with varying numbers of diffuser units, $M$. **(b)** Output PCC values of diffractive networks trained with different numbers of diffuser units, evaluated using known, new, and no diffusers. **(c)** Representative object reconstructions showing the effect of increasing $M$ on the output image quality.

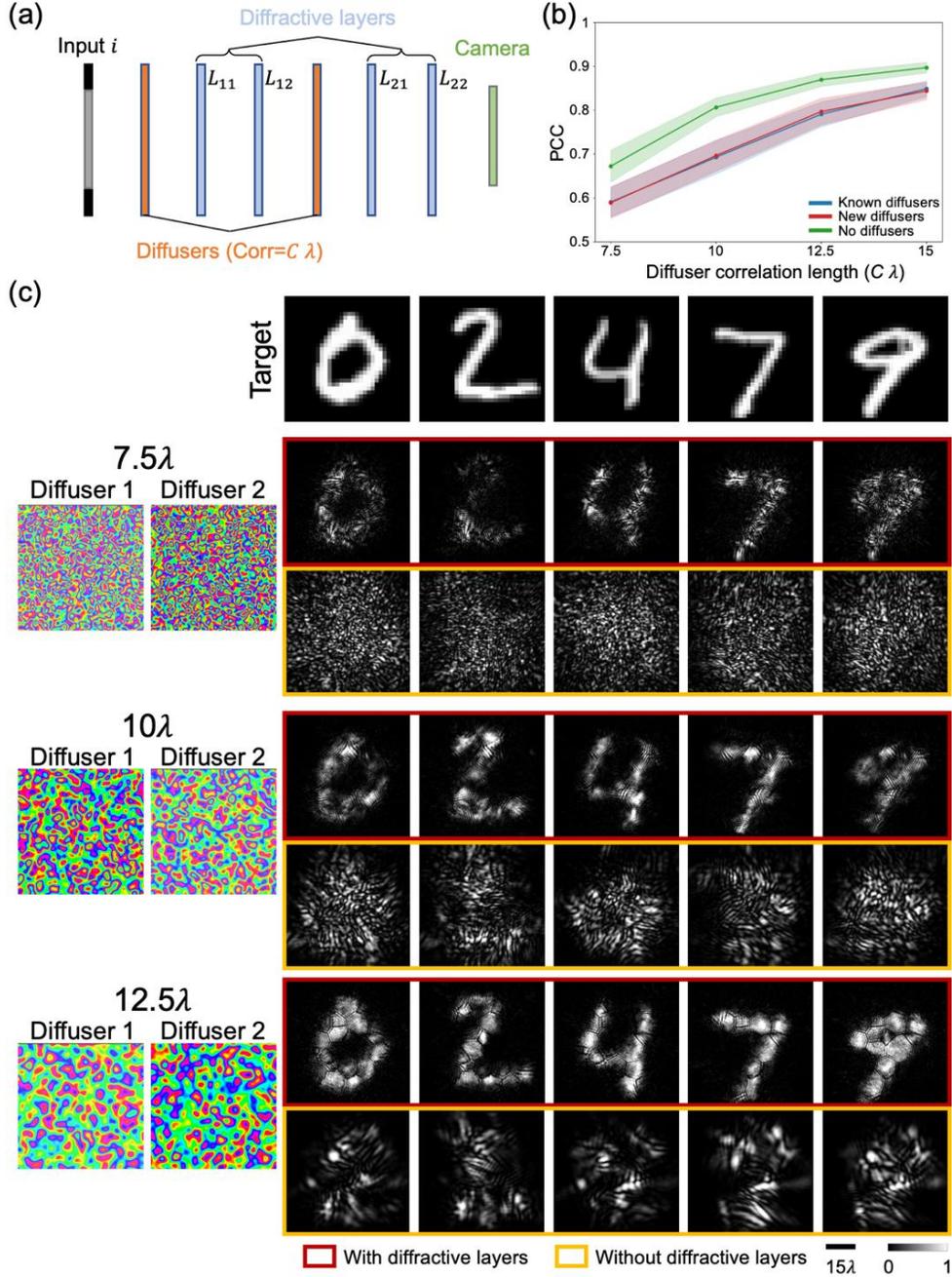

**Figure 5. Impact of the diffuser correlation length on the imaging performance. (a)** Schematic of the interleaved diffractive network with two diffuser units, where the diffuser correlation length was varied as $C$ (in units of $\lambda$). **(b)** The output PCC values of diffractive networks trained with different diffuser correlation lengths $C$, evaluated using known, new, and no diffusers. **(c)** Representative reconstruction results for varying diffuser correlation lengths ($C = 7.5\lambda, 10\lambda, 12.5\lambda$). Red boxes denote systems with optimized diffractive layers, while yellow boxes show the outputs without diffractive layers.

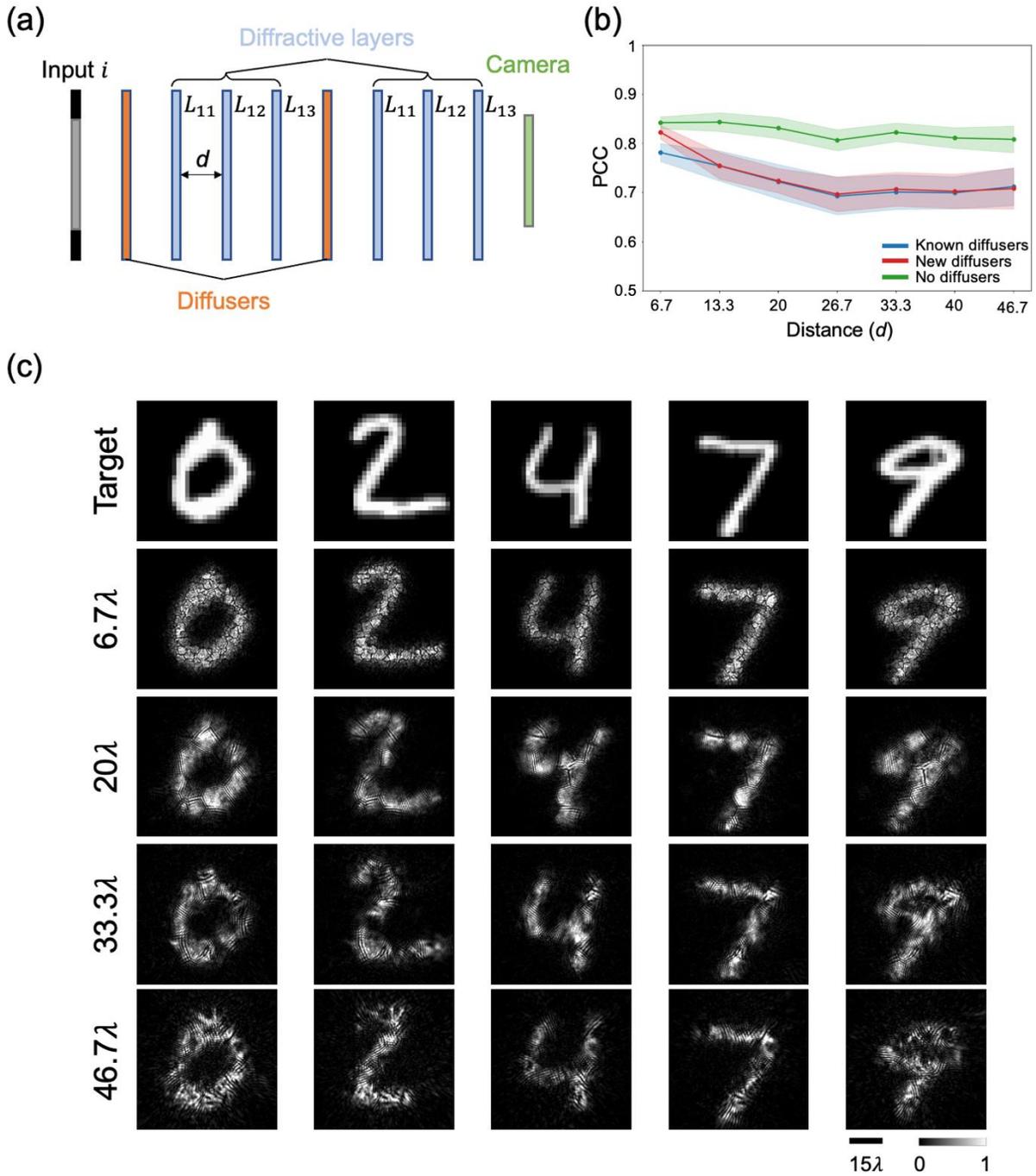

**Figure 6. Impact of the axial separation between the diffractive layers on the imaging performance. (a)** Schematic of the interleaved diffractive network with $M = 2$ diffuser units and $K = 3$ diffractive layers, where the axial distance ($d$) between consecutive planes was varied. **(b)** PCC values of the diffractive networks trained with different inter-plane distances, evaluated using known, new, and no diffusers. **(c)** Reconstruction results of varying $d$ ($6.7\lambda - 46.7\lambda$).

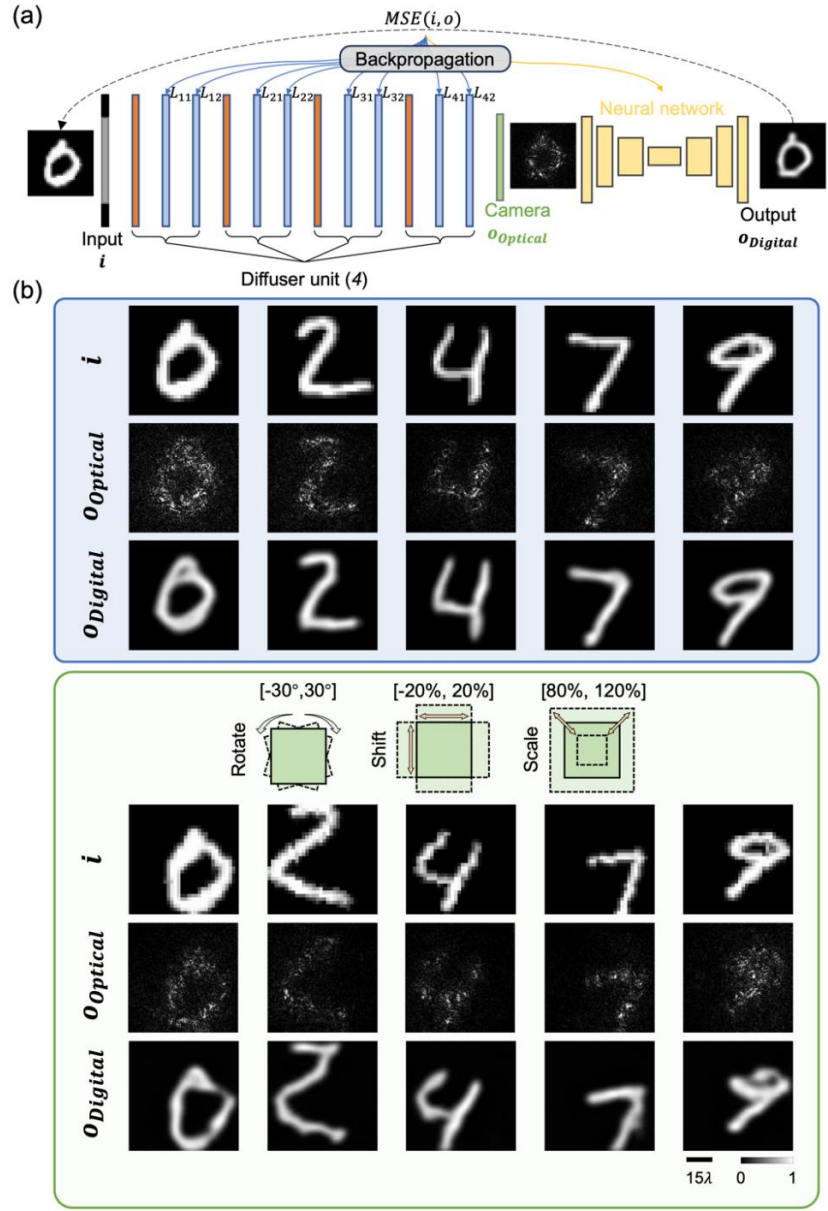

**Figure 7. Hybrid optical-digital interleaved diffractive network for optical information transfer. (a)** Schematic of the hybrid system. An interleaved diffractive network performs physical pre-processing, forming an intermediate image on a camera, followed by a neural network that digitally reconstructs the final high-fidelity output. Both the optical and digital components are jointly optimized in an end-to-end manner. **(b)** Comparison of the all-optical and the hybrid outputs through four diffuser units ($M = 4$); also see **Movie S2**. The hybrid network significantly improves the image reconstruction fidelity for complex scattering conditions (blue box). The green box illustrates the robustness of the hybrid system to input perturbations, including randomly rotations, lateral shifts, and scaling of the input objects.

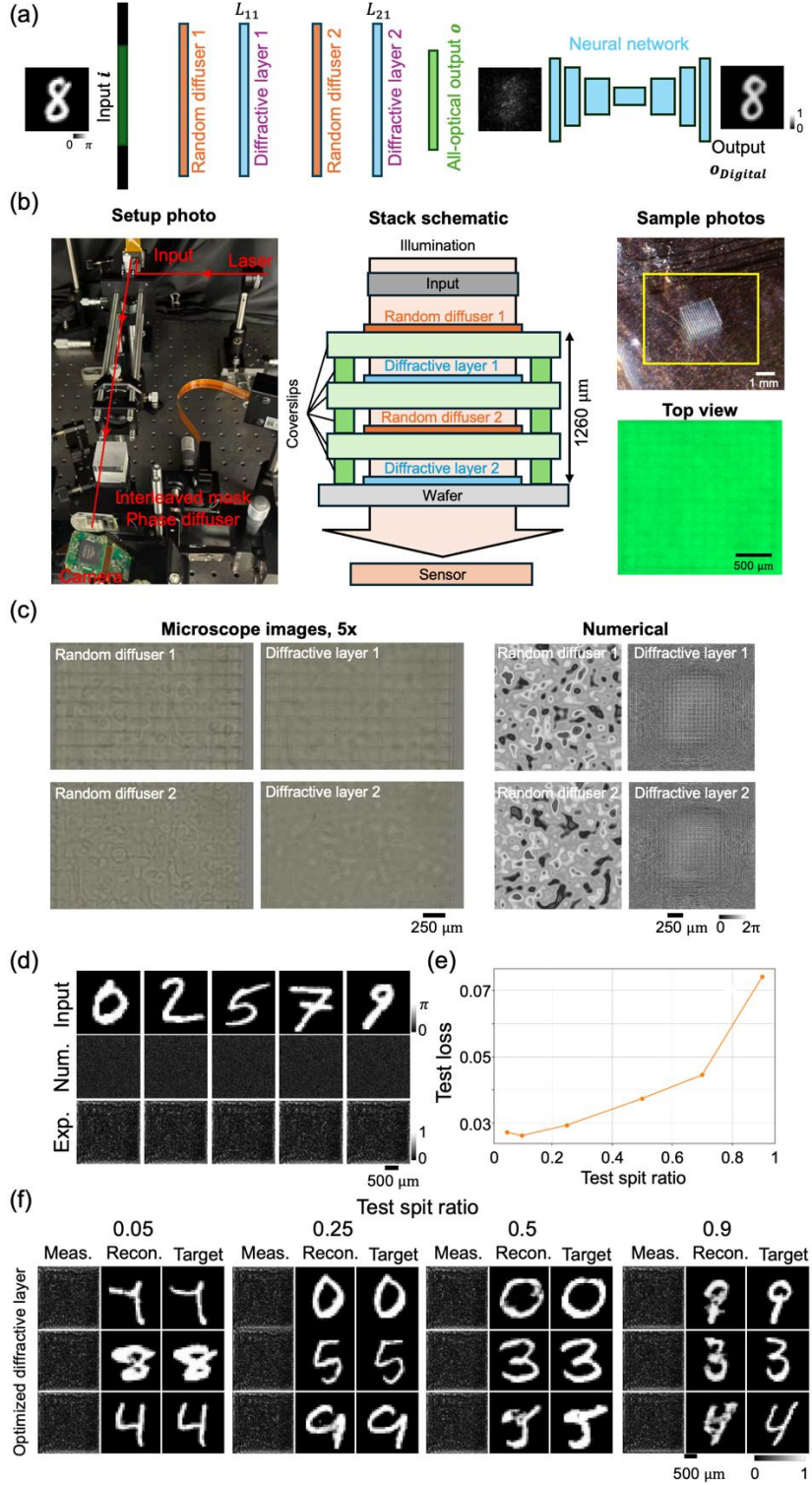

**Figure 8. Experimental testing of an interleaved diffractive network. (a)** Schematic illustration of the hybrid optical-digital system used for training. Two optimized diffractive layers were interleaved with two random phase diffusers (never used during in silico training). The optical

output intensity was further processed by a digital neural network backend. **(b)** A photograph of the experimental setup and a schematic of the 3D fabricated layer stack. The fabricated four-layer stack (two random phase diffusers interleaved with two spatially optimized diffractive layers) was illuminated by the input object displayed on an SLM, and the output intensity was recorded by a CMOS camera for digital reconstruction. Sample photographs and a top-view microscopic image of the 3D fabricated structures are also shown. **(c)** Microscopic images (5 × objective) of the fabricated random phase diffusers and optimized diffractive layers, together with their corresponding numerical phase profiles are shown. **(d)** Comparison between numerical output and experimental measurements at the CMOS image sensor. Top row: input digits. Middle row: numerically predicted output intensity. Bottom row: experimentally measured CMOS intensity distributions. **(e)** Test loss as a function of the test split ratio for the optimized interleaved diffractive masks. **(f)** Representative experimental measurements (Meas.), digital reconstructions (Recon.), and ground truth targets (Target) for different test split ratios. In the experiments, the input phase images were defined over the range $[0, \pi]$, while the corresponding target images were normalized to $[0,1]$.